\documentclass[twocolumn,showpacs,floatfix,aps]{revtex4}
\usepackage{graphicx}
\usepackage{dcolumn}
\usepackage{bm}

\begin{document}


\title{Two-particle localization and antiresonance \\ in disordered spin
 and qubit chains}

\author{L. F. Santos}
\email{santos@pa.msu.edu}
\author{M. I. Dykman}
\email{dykman@pa.msu.edu}
\affiliation{Department of Physics and Astronomy
and the Institute for Quantum Sciences,
Michigan State University, East Lansing, Michigan 48824}

\date{\today}

\begin{abstract}
We show that, in a system with defects, two-particle states may
experience destructive quantum interference, or antiresonance. It
prevents an excitation localized on a defect from decaying even
where the decay is allowed by energy conservation. The system studied
is a qubit chain or an equivalent spin chain with an anisotropic ($XXZ$)
exchange coupling in a magnetic field. The chain has a defect with an
excess on-site  energy. It corresponds to a qubit with the
level spacing different from other qubits.  We show that, because of
the interaction between excitations, a single defect may lead to
multiple localized states. The energy spectra and localization lengths
are found for two-excitation states. The localization of
excitations facilitates the operation of a quantum
computer. Analytical results for strongly anisotropic coupling are
confirmed by numerical studies.
\end{abstract}


\maketitle

\section{Introduction}

One of the most important potential applications of quantum computers
(QC's) is studies of quantum many-body effects. It is particularly
interesting to find new many-body effects in condensed-matter systems
that could be easily simulated on a QC. In the present paper we
discuss one such effect: antiresonance, or destructive quantum
interference between two-particle excitations in a system with
defects. We also study interaction-induced two-particle localization on
a defect and discuss implications of the results for quantum
computing.

The basic elements of a QC, qubits, are two-state systems. They are
naturally modeled by spin-1/2 particles. In many suggested
realizations of QC's, the qubit-qubit interaction is ``on'' all the time
\cite{recent_reviews}. In terms of spins, it corresponds to exchange
interaction. The dynamics of such QC's and spin systems in solids have
many important similar aspects that can be studied together.

In most proposed QC's the energy difference between the qubit states is
large compared to the qubit-qubit interaction. This corresponds to a
system of spins in a strong external magnetic field. However, in
contrast to ideal spin systems, level spacings of different qubits
can be different. A major advantageous feature of QC's is that the qubit
energies can be often individually controlled
\cite{Makhlin99,JJ-all,mark}. This corresponds to controllable
disorder of a spin system, and it allows one to use QC's for studying a
fundamentally important problem of how the spin-spin interaction
affects spin dynamics in the presence of disorder.


Several models of QC's where the interqubit interaction is permanently
``on'' are currently studied. In these models the effective spin-spin
interaction is usually strongly anisotropic. It varies from the
essentially Ising coupling $\sigma_n^z\sigma_m^z$ in nuclear
magnetic resonance and some
other systems
\cite{Chuang9802,Yamamoto02,Cory00,Piermarocchi02,Demille02} ($n,m$
enumerate qubits, and $z$ is the direction of the magnetic field) to
the $XY$-type (i.e., $\sigma_n^x\sigma_m^x + \sigma_n^y\sigma_m^y$) or
the $XXZ$-type (i.e., $\Delta_{nm}\sigma_n^z\sigma_m^z+
\sigma_n^x\sigma_m^x + \sigma_n^y\sigma_m^y$) coupling in some
Josephson-junction based systems \cite{Makhlin99,JJ-all}.

The Ising coupling describes the system in the case where the
transition frequencies of different qubits are strongly
different. Then $\sigma_n^z\sigma_m^z$ is the only part of the
interaction that slowly oscillates in time, in the Heisenberg
representation, and therefore is not averaged out. If the qubit
frequencies are close to each other, the terms $\sigma_n^x\sigma_m^x +
\sigma_n^y\sigma_m^y$ become smooth functions of time as well. They
lead to resonant excitation hopping between qubits.  In a multiqubit
system with close frequencies, both Ising and $XY$ interactions are
present in the general case, but their strengths may be different
\cite{Silvestrov01,Kaminsky-Lloyd02}. In this sense the $XXZ$
coupling is most general, at least for qubits with high transition
frequencies.

The interqubit interaction often rapidly falls off with the distance
and can be approximated by nearest neighbor coupling. Many important
results on anisotropic spin systems with such coupling have been
obtained using the Bethe ansatz. Initially the emphasis was placed on
systems without defects \cite{Bethe} or with defects on the edge of a
spin chain \cite{Baxter}. More recently these studies have been
extended to systems with defects that are described by integrable
Hamiltonians \cite{integrable}. However, the problem of a spin chain
with several coupled excitations and with defects of a general type
has not been solved.

In this paper we investigate interacting 
excitations in an anisotropic spin system
with defects. We show that the excitation localized on a defect does
not decay even where the decay is allowed by energy conservation.  We
also find that, in addition to a single-particle excitation, a defect
leads to the onset of two types of localized two-particle excitations.

The analysis is done for a system with the $XXZ$ coupling. The coupling
anisotropy is assumed to be strong, as in the case of a QC based on
electrons on helium, for example \cite{mark}. The ground state of the
system corresponds to all spins pointing in the same direction
(downwards, for concreteness).  A single-particle excitation
corresponds to one qubit being excited, or one spin being flipped. If
the qubit energies are tuned in resonance with each other, a QC
behaves as an ideal spin system with no disorder. A single-particle
excitation is then magnon-type, it freely propagates through the
system.

In the opposite case where the qubit energies are tuned far away from
each other (as for diagonal disorder in tight-binding models), all
single-particle excitations are localized. If the excitation density
is high, the interaction between them may affect their localization,
leading to quantum chaos, cf. Refs.
\onlinecite{Silvestrov01,Shepelyansky00,Izrailev01,Lloyd02}.
Understanding the interplay between interaction and disorder is a
prerequisite for building a QC. We will consider the case where the
excitation density is low, yet the interaction is important. In
particular, excitations may form bound pairs (but the pair density is
small).

One of the important questions is whether the interaction leads to
delocalization of excitations. More specifically, consider an
excitation, which is localized on a defect in the absence of other
excitations. We now create an extended magnon-type excitation (a
propagating wave), that can be scattered off the localized one.
The problem is whether this will
cause the excitation to move away from the defect. We show below that,
due to unexpected destructive quantum interference, the scattering
does not lead to delocalization.

\subsection{Model and preview}
We consider a one-dimensional array of qubits which models a spin-1/2
chain. For nearest-neighbor coupling, the Hamiltonian is
\begin{eqnarray}
\label{hamiltonian}
&&H =  \frac{1}{2}\sum_n \varepsilon^{(n)} \sigma_{n}^{z} +
\frac{1}{4}\sum_n \sum_{i=x,y,z}J_{ii}\sigma_n^i\sigma_{n+1}^i\\
&&J_{xx}=J_{yy}=J,\;J_{zz}=J\Delta\nonumber .
\end{eqnarray}
Here, $\sigma_n^i$ are the Pauli matrices and $\hbar =1$. The
parameter $J$ characterizes the strength of the exchange coupling, and
$\Delta$ determines the coupling anisotropy. We assume that
$|\Delta|\gg 1$; for a QC based on electrons on helium, $|\Delta|$
lies between $20$ and $8$, for typical parameter values
\cite{mark}.

We will consider effects due to a single defect. Respectively, all
on-site spin-flip energies $\varepsilon^{(n)}$ are assumed to be the
same except for the site $n=n_0$ where the defect is located, that is,
\begin{equation}
\label{chain_energies}
\varepsilon^{(n)}= \varepsilon + g\delta_{n,n_0} .
\end{equation}

\begin{figure}[ht]
\includegraphics[width=3.2in]{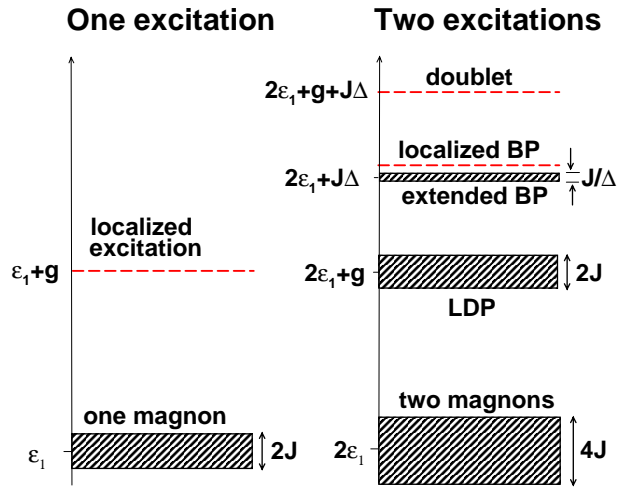}
\caption{Left panel: the one-excitation energy spectrum in an infinite
spin chain with a defect.  The energies of extended states (magnons)
form a band of width $J$ centered at ${\varepsilon}_1$. The dashed
line indicates the energy of the excitation localized on the
defect. Right panel: the two-excitation energy spectrum. The band
$2({\varepsilon}_1\pm J)$ is formed by uncoupled magnons. The band
centered at $2{\varepsilon}_1+g$ is formed by localized-delocalized
pairs (LDP's) in which one excitation is localized on the defect and
the other is in an extended state. The narrow band centered at
$2{\varepsilon}_1 + J\Delta$ is formed by propagating bound pairs (BP's)
of excitations.  The dashed lines show the energies of the
states where both excitations are bound to the defect.}
\label{fig:energies}
\end{figure}

In order to formulate the problem of interaction-induced decay of
localized excitations, we preview in Fig.~\ref{fig:energies} a part of
the results on the energy spectrum of the system.  In the absence of
the defect, the energies of single-spin excitations (magnons) lie
within the band ${\varepsilon}_1 \pm J$, where ${\varepsilon}_1
=\varepsilon-J\Delta $ (the energy is counted off from the 
ground-state energy).  The defect has a spin-flip energy that differs by $g$
(a qubit with a transition frequency different from that of other
qubits).  It leads to a localized single-spin excitation with no
threshold in $g$, for an infinite chain. The energy of the localized
state is shown by a dashed line on the left panel of
Fig.~\ref{fig:energies}.

We now discuss excitations that correspond to two flipped spins. A
defect-free $XXZ$ system has a two-magnon band of independently
propagating noninteracting magnons. However, the anisotropy of the
exchange coupling leads also to the onset of bound pairs (BP's) of
excitations. The BP band is much narrower than the two-magnon band and
is separated from it by a comparatively large energy difference
$J\Delta$, see the right panel of Fig.~\ref{fig:energies}. In the
presence of a defect, there are two-excitation states with one
excitation localized on the defect and the other being in an extended
state. We call them localized-delocalized pairs (LDP's).  An interplay
between disorder and interaction may lead to new types of states where
both excitations are localized near the defect. Their energies are
shown in the right panel of Fig.~\ref{fig:energies} by dashed lines.

A localized one-spin excitation cannot decay by emitting a magnon, by
energy conservation. But  it might experience an
induced decay when a magnon is inelastically scattered off the excited
defect into an extended many-spin state. Magnon-induced decay is
allowed by energy conservation when the total energy of the localized
one-spin excitation and the magnon coincides with the energy of
another two-particle state.  In the $XXZ$ model the total number of
excitations (flipped spins) is conserved, and therefore decay is only
possible into extended states of two bound magnons. In other words, it
may only happen when the LDP band overlaps with the BP band in
Fig.~\ref{fig:energies}.

Decay into BP states may occur directly or via the two-excitation
state located next to the defect. The amplitudes of the corresponding
transitions turn out to be nearly equal and opposite in sign. As a
result of this quantum interference, even though the band of bound
magnons is narrow and has high density of states, the LDP to BP
scattering does not happen, i.e., the excitation on the defect is not
delocalized. The BP to LDP scattering does not happen either, i.e., a
localized excitation is not created as a result of BP decay.

In Sec.~II below and in the Appendix we briefly analyze localization of
one excitation in a finite chain with a defect, for different boundary
conditions. In Sec.~III we discuss the two-excitation states localized
near a defect. In Sec.~IV we consider the resonant situation where the
energy band of extended bound pair states is within the band of
energies of the flipped defect spin plus a magnon, i.e., where the BP
band overlaps with the LDP band in Fig.~\ref{fig:energies}.  We find
that the localized excitation remains on the defect site in this
case. Analytical results for a chain with strong anisotropy $|\Delta|$
are compared with numerical calculations. Section V contains concluding
remarks.

\section{One excitation: localized and extended states}

In order to set the scene for the analysis of the two-excitation case,
in this section and in the Appendix we briefly discuss the well-known case
of one excitation (flipped spin) in an $XXZ$ spin chain with a defect
\cite{Economou} and the role of boundary conditions.  The Hamiltonian
of the chain with the defect on site $n_0$ has the form
(\ref{hamiltonian}), (\ref{chain_energies}). We assume that the
excitation energy $\varepsilon$ largely exceeds both the coupling
constant $|J|$ and the energy excess on the defect site $|g|$.
In this case the ground-state of the system corresponds to all spins
being parallel, with $\langle\sigma_n^z\rangle=-1$
irrespective of the signs of $J,g$, and $\Delta$.

Without a defect, one-spin excitations are magnons. They freely
propagate throughout the chain. The term in the Hamiltonian
(\ref{hamiltonian}) responsible for one-excitation hopping is
$H^{(t)}=\sum\nolimits_n H^{(t)}_n$, with
\begin{eqnarray}
\label{transfer_one}
&H^{(t)}_n = {1\over 4}J \sum_{i=x,y} \sigma_n^i\sigma_{n+1}^i \equiv {1\over
8}J (t_n^{(l)} + t_n^{(r)}),\\
&t_n^{(r)}= [t_{n}^{(l)}]^{\dagger}=
\sigma_{n+1}^{+} \sigma_{n}^{-}.\nonumber
\end{eqnarray}
The operators $t_{n}^{(r)}$ and $t_{n}^{(l)}$ cause excitation
shifts $n\to n+1$ and $n+1\to n$, respectively.

A defect leads to magnon scattering and to the onset of localized
states.  Both propagation of excitations and their localization are
interesting for quantum computing. Coherent excitation transitions allow
one to have a QC geometry where remote ``working'' qubits are
connected by chains of ``auxiliary'' qubits, which form ``transmission
lines'' \cite{mark}. Localization, on the other hand, allows one to
perform single-qubit operations on targeted qubit.

A QC makes it possible to model spin chains with different boundary
conditions. The simplest models are an open spin chain with free
boundaries, which is mimicked by a finite-length array of qubits (for
electrons on helium, it can be implemented using an array of equally
spaced electrodes, cf. Ref.~\onlinecite{Goodkind01}) or a periodic
chain, which can be mimicked by a ring of qubits.

An open $N$-spin chain is described by the Hamiltonian
(\ref{hamiltonian}), where the first sum runs over $n=1,\ldots,N$ and
the second sum runs over $n=1,\ldots,N-1$ (the edge spins have
neighbors only inside the chain). In what follows, we count energy off
the ground-state energy ${E}_{0} = -(N\varepsilon+g)/2 + (N-1) J
\Delta /4 $, i.e., we replace in Eq.~(\ref{hamiltonian}) $H\to H-E_0$.

The eigenfunctions of $H$ in  the case of one
excitation can be written as \cite{Economou}

\begin{equation}
\psi_1 = \sum_{n=1}^{N} a(n) \phi (n)
\label{one_exc} ,
\end{equation}
where $\phi (n) $ corresponds to the spin on site $n$ being up
and all other spins being down.  The Schr\"odinger equation for
$a(n) $ has the form

\begin{eqnarray}
&& \left( {\varepsilon}_{1} + g \delta_{n,n_0}
+\frac{J\Delta }{2} \delta_{n,1}
+\frac{J\Delta }{2} \delta_{n,N} \right) a(n)
\nonumber \\
        &&+
\frac{J }{2} [a(n-1) + a(n+1)]=E_1a(n),
\label{eq_a}
\end{eqnarray}
where ${\varepsilon}_{1} = \varepsilon -J\Delta$ is the energy of a
flipped spin in an ideal infinite chain in the absence of excitation
hopping and $E_1$ is the one-excitation energy eigenvalue. For an
open $N$-spin chain we set $a(0)= a(N+1)=0$ in
Eq.~(\ref{eq_a}).

The Hamiltonian of a closed $N$-spin chain has the form
(\ref{hamiltonian}), where both sums over $n$ go from 1 to $N$ and the
site $N+1$ coincides with the site 1. Here, the defect
location $n_0$ can be chosen arbitrarily.
The wave function can be sought in the
form (\ref{one_exc}). The Schr\"odinger equation then has the form
(\ref{eq_a}), except that there are no terms proportional to $\delta_{n,1},\,
\delta_{n,N}$ from the end points of the chain. It has to be solved
with the boundary condition $a(n+N)=a(n)$.

For an open chain, the solution of the Schr\"odinger equation
(\ref{eq_a})  can be sought in the form of plane
waves propagating between the chain boundaries and the defect,
\begin{equation}
\label{left_right}
a (n) =
  C_{l,r}e^{i\theta n} + C'_{l,r} e^{- i\theta n}, \quad |n-n_0|\geq 1 .
\end{equation}
The subscripts $l$ and $r$ refer to the coefficients for the waves to
the left ($n<n_0$) and to the right ($n>n_0$) from the defect. The
interrelations between these coefficients and the coefficient $a(n_0)$
follow from the boundary conditions and from matching the solutions at
$n_0$. They are given by Eqs.~(\ref{bound_cond}) and
(\ref{boundary_defect}).

For a closed chain, on the other hand, the solution can be sought
in the form
\begin{eqnarray}
\label{right}
a(n) &=& Ce^{i\theta n} + C'e^{-i\theta n}, \quad n_0<n\leq
N, \\
a(n) &=& Ce^{i\theta (n+N)} + C'e^{-i\theta (n+N)}, \quad 1\leq n
< n_0.\nonumber
\end{eqnarray}

The energy $E_1$ as a function of $\theta$ can be obtained by
substituting Eqs.~(\ref{left_right}) and (\ref{right}) into
Eq.~(\ref{eq_a}). Both for the open and closed chains it has the form
\begin{equation}
E_1 = {\varepsilon}_{1} +
J \cos \theta .
\label{energy_one}
\end{equation}

The eigenfunctions (\ref{left_right}), (\ref{right}) with real
$\theta$ correspond to sinusoidal waves (extended states). From
(\ref{energy_one}), their energies lie within the band
${\varepsilon}_{1} \pm J$. In contrast, localized states have complex
$\theta$, and their energies lie outside this band. The corresponding
solutions for both types of chains are discussed in the Appendix. In a
sufficiently long chain, there is always one localized one-excitation
state on a defect. Its energy is given by Eq.~(\ref{E_d}). In the case
of an open chain with the coupling anisotropy parameter $|\Delta|
> 1$, there are also localized states on the chain boundaries. Their
energy is given by Eq.~(\ref{E_s}).

\section{Two excitations: unbound, bound, and localized states}

A spin chain with a defect displays rich behavior in the presence
of two excitations. It is determined by the interplay between
disorder and inter-excitation coupling.  Solutions of the
two-excitation problem have been obtained in the case of a
disorder potential of several special types, where the system is
integrable \cite{integrable}. Here we will study the presumably
nonintegrable but physically interesting problem where the
on-site energy of the defect differs from that of the host sites.

The system is described by Hamiltonian (\ref{hamiltonian}). 
In order to concentrate on the effects of
disorder rather than boundaries, we will consider a closed chain of
length $N$. We will also assume that the anisotropy is strong,
$|\Delta| \gg 1$.

The wave function of a chain with two excitations is given by a linear
superposition
\begin{equation}
\label{two_exc}
\psi_2  = \sum_{n<m} a(n,m) \phi (n,m),
\end{equation}
where $\phi(n,m)$ is the state where spins on the sites $n$ and $m$
are pointing upward, whereas all other spins are pointing downward. In
a periodic chain, the sites with numbers that differ by $N$ are
identical, therefore we have $a(n,m) = a(m, n+N)$.

From Eq.~(\ref{hamiltonian}), the Schr\"odinger equation for the
coefficients $a(n,m)$  is
\begin{widetext}
\begin{eqnarray}
\label{a(n,m)}
&&\left( 2{\varepsilon}_{1} + g \delta_{n,n_0}+ g \delta_{m,n_0}
+J\Delta  \delta_{m,n+1}\right)a(n,m)
\nonumber \\
       && +\frac{1}{2}J \bigl[a(n-1,m) + [a(n+1,m) +
a(n,m-1)](1-\delta_{m,n+1}) + a(n,m+1)\bigr] =E_2a(n,m).
\end{eqnarray}
\end{widetext}
Here, $E_2$ is the energy of a two-excitation state [${\varepsilon}_1 =
\varepsilon - J\Delta$ is the on-site one-excitation energy,
cf. Eq.~(\ref{eq_a})]. As before, we assume that the defect is located
on site $n_0$.

\subsection{An ideal chain}

In the absence of a defect the system is integrable. The solution of
Eq.~(\ref{a(n,m)}) can be found using the Bethe ansatz \cite{Bethe},
\begin{equation}
\label{two_bethe_ideal}
a(n,m) = C e^{i (\theta_1 n+ \theta_2 m)} +
C' e^{i (\theta_2 n +  \theta_1 m)}.
\end{equation}
The energy of the state with given $\theta_1,\theta_2$ is obtained by
substituting Eq.~(\ref{two_bethe_ideal}) into Eq.~(\ref{a(n,m)})
written for $m>n+1$. This gives
\begin{equation}
\label{E2}
E_2 = 2{\varepsilon}_{1} + J (\cos \theta_1 + \cos \theta_2).
\end{equation}

By requiring that the ansatz (\ref{two_bethe_ideal}) apply also for
$m=n+1$, we obtain an interrelation between the coefficients $C$ and $C'$,
\begin{equation}
\label{CC'}
\frac{C}{C'} =
-\frac{1 - 2 \Delta  e^{i \theta_1 } + e^{i (\theta_1 + \theta_2)}}
{1 - 2 \Delta  e^{i \theta_2 } + e^{i (\theta_1 + \theta_2)}}.
\end{equation}
With account taken of normalization, Eqs.~(\ref{two_bethe_ideal}) and
(\ref{CC'}) fully determine the wave function.
The states with real $\theta_{1,2}$ form a two-magnon band with width
$4|J|$, as seen from the dispersion relation (\ref{E2}). The magnons
are not bound to each other and propagate independently.

For $|\Delta| > 1$, Eq.~(\ref{CC'}) also has a solution $C=0$, which
gives a complex phase $\theta_{2}= \theta_1^* = {1\over 2}\theta
-i\kappa$ with $\kappa > 0$. This solution corresponds to the wave
function $a(n,m)\propto \exp[i\theta(n+m)/2 - \kappa (m-n)]$;
we have $m>n$. From
Eqs.~(\ref{two_bethe_ideal})--(\ref{CC'}) we obtain
\begin{eqnarray}
&&e^{-\kappa}=\Delta^{-1}\cos(\theta/2),\quad E_{BP} = E_{BP}^{(0)}+
{J\over 2\Delta}\cos \theta,\nonumber\\
&&
E_{BP}^{(0)}= 2{\varepsilon}_1 + J\Delta + {J\over 2\Delta}.
\label{bound}
\end{eqnarray}

Eq.~(\ref{bound}) describes a bound pair of excitations. Such a
pair can freely propagate along the chain. The wave function is
maximal when the excitations are on neighboring sites.  The size of
the BP, i.e. the typical distance between the excitations, is
determined by the reciprocal decrement $\kappa^{-1}$, and ultimately
by the anisotropy parameter $\Delta$. For large $|\Delta|$, the
excitations in a BP are nearly completely bound to nearest sites. Then
the coefficients $a(n,m)\propto \delta_{n+1,m}$, to the lowest order
in $|\Delta|^{-1}$.

The distance between the centers of the BP band and the two-magnon
band $E_{BP}^{(0)}-2\varepsilon_1$ is given approximately by the BP
binding energy $J\Delta$. The width of the BP band $|J/\Delta|$ is
parametrically smaller than the width of the two-magnon band $4|J|$,
see Fig.~\ref{fig:energies}.

For nearest-neighbor coupling and for large $|\Delta|$, transport of
bound pairs can be visualized as occurring via an intermediate
step. First, one of the excitations in the pair makes a virtual
transition to the neighboring empty site, and as a result the parallel
spins in the pair are separated by one site. The corresponding state
differs in energy by $J\Delta$ from the bound-pair state. At the next
step the second spin can move next to the first, and then the whole
pair moves by one site. From perturbation theory, the bandwidth should
be $J^2/J\Delta\equiv J/\Delta$, which agrees with Eq.~(\ref{bound}).

The above arguments can be made quantitative by introducing an
effective Hamiltonian $\tilde H^{(t)} = \sum_{n=1}^{N} \tilde
H_{n}^{(t)} $ of BP's. It is obtained from the Hamiltonian
$H^{(t)}$ (\ref{transfer_one}) in the second order of perturbation
theory in the one-excitation hopping constant $J$,
\begin{eqnarray}
\label{transfer_bound}
\tilde H_n^{(t)}&=&\frac{J }{64 \Delta  }
[t^{(r)}_{n-1} t^{(l)}_{n-1} + t^{(l)}_{n+1} t^{(r)}_{n+1}]
\\
&+& \frac{J }{64 \Delta  }
[t^{(l)}_{n} t^{(l)}_{n-1} + t^{(r)}_{n} t^{(r)}_{n+1}].
\nonumber
\end{eqnarray}
The operators $t^{(r,l)}_n$ of excitation hopping to the right or
left are given by Eq.~(\ref{transfer_one}). The first pair of terms in
$\tilde H_n^{(t)}$ [Eq.~(\ref{transfer_bound})] describes virtual
transitions in which a BP dissociates and then recombines on the same
site. This leads to a shift of the on-site energy level of the BP by
$J/2\Delta$. The second pair of terms describes the motion of a BP as
a whole to the left or to the right.

The action of the Hamiltonian $\tilde H^{(t)}$ on the wave function
$a(n,n+1)$ is given by
\begin{eqnarray}
\label{transferBP}
&&\tilde H^{(t)}a(n,n+1)=\\
&&{J\over 2\Delta}a(n,n+1) + {J\over
4\Delta}[a(n+1,n+2) + a(n-1,n)]. \nonumber
\end{eqnarray}
The Schr\"odinger equation for a BP is given by the sum of
the diagonal part (the first term) of Eq.~(\ref{a(n,m)}) with $m=n+1$
and the right-hand side of Eq.~(\ref{transferBP}).  In this
approximation a BP eigenfunction is $a(n,m)=\delta_{n+1,m}\exp(i\theta
n)$, and the dispersion law is of the form (\ref{bound}).

\subsection{Localized states in a chain with a defect}

We now consider excitations in the presence of a defect. In this section we
assume that the defect excess energy $g$ is such that
\begin{equation}
\label{nonresonant_cond}
|g|\gg |J|, \quad |J\Delta-g| \gg |J| .
\end{equation}
The first inequality guarantees that the localization length of an
excitation on the defect is small [its inverse Im~$\theta_d\approx
\ln 2|g/J| \gg 1$ cf. Eq.~(\ref{one_localized})].

The second condition in Eq.~(\ref{nonresonant_cond}) can be understood by
noticing that,
in a chain with a defect, there is a two-excitation state where one
excitation is localized on the defect whereas the other is in an
extended magnon-type state. These excitations are not bound
together. The energy of such an unbound localized-delocalized pair
(LDP) should differ from the energy of unbound pair of magnons by
$\approx g$, and from the energy of a bound pair of magnons
(\ref{bound}) by $\approx J\Delta - g$. In this section we consider
the case where both these energy differences largely exceed the magnon
bandwidth $J$ (the case where $|J\Delta - g|\alt |J|$ will be
discussed in the following section).

\subsubsection{Unbound localized-delocalized pairs (LDP's)}

In the neglect of excitation hopping, the energy of an excitation
pair where one excitation is far from the defect ($|n-n_0|\gg 1$) and
the other is localized on the defect is $2{\varepsilon}_1 + g$. [This
can be seen from Eq.~(\ref{a(n,m)}) for the coefficients $a(n,m)$ in
which off-diagonal terms are disregarded.] At the same time, if one
excitation is on the defect and the other is on the neighboring site
$n_0\pm 1$, this energy becomes $2{\varepsilon}_1 + g + J\Delta$. The
energy difference $J\Delta$ largely exceeds the characteristic
bandwidth $J$. Therefore, if the excitation was initially far from the
defect, it will be reflected before it reaches the site $n_0\pm 1$.

The above arguments suggest to seek the solution for the wave function
of an LDP in a periodic chain in the form
\begin{equation}
\label{LDP_wave}
a(n_0,m)=Ce^{i\theta m} + C'e^{-i\theta m} ,
\end{equation}
with the boundary condition $a(n_0, n_0 +1)= a(n_0,n_0+N-1) = 0$. This
boundary condition and the form of the solution are similar to what
was used in the problem of one excitation in an open chain.

From Eq.~(\ref{a(n,m)}), the energy of an LDP is
\begin{equation}
\label{energy_LDP}
E_{LDP}\approx 2{\varepsilon}_1 + g + J\cos\theta .
\end{equation}
The wave number $\theta$ takes on $N-3$ values $\pi k/(N-2)$, with
$k=1,\ldots, N-3$.
As expected, the bandwidth of the LDP is $2|J|$, as in the case of
one-excitation band in an ideal chain.

We have compared Eq.~(\ref{energy_LDP}) with numerical results
obtained by direct solution of the eigenvalue problem
(\ref{a(n,m)}). For $N=100, \Delta = 20, g/J=10$ we obtained excellent
agreement once we took into account that the energy levels
(\ref{energy_LDP}) are additionally shifted by $J^2/2g$. This shift
can be readily obtained from Eq.~(\ref{a(n,m)}) as the second-order
correction (in $J$) to the energy $E_d$ of the excitation localized on
the defect. It follows also from Eq.~(\ref{E_d}) for $|g|\gg |J|$.

We note that the result is trivially generalized to the case of a
finite but small density of magnons. The wave function $a(n_0,
m_1,m_2,\ldots,m_M)$ of $M$ uncoupled magnons and an excitation on the
site $n_0$ is given by a sum of the appropriately weighted
permutations of $\exp(\pm i\theta_1m_1 \pm i\theta_2m_2 \ldots \pm
i\theta_Mm_M)$ over the site numbers $m_i$ (these numbers can be
arranged so that $m_1<\ldots <m_k<n_0<m_{k+1}<\ldots < m_M$), with
real $\theta_i$. The weighting factors for large $|\Delta|$ are found
from the boundary conditions and the condition that $a(n_0,
m_1,m_2,\ldots)=0$ whenever any two numbers $m_i, m_{i+1}$ differ by
1. When the ratio of the number of excitations $M$ to the chain length
$N$ is small, the energy is just a sum of the energies of uncoupled
magnons and the localized excitation, i.e., it is
$(M+1){\varepsilon}_1 + g + J\sum\nolimits_{i=1}^M\cos\theta_i$. For
small $M/N$, scattering of a magnon by the excitation on a defect
occurs as if there were no other magnons, i.e., the probability of a
three-particle collision is negligibly small.

\subsubsection{Bound pairs localized on the defect: the doublet}

For large $|\Delta|,|g/J|$, a bound pair of neighboring excitations
should be strongly localized when one of the excitations is on the
defect site. Indeed, if we disregard intersite excitation hopping,
the energy of a BP sitting on the defect is
$E_D^{(0)}=2{\varepsilon}_1+ g+ J\Delta $. It differs significantly
from the energy of freely propagating BP's (\ref{bound}), causing
localization.

The major effect of the excitation hopping is that the pair can make
resonant transitions between the sites $(n_0, n_0+1)$ and $(n_0-1,
n_0)$. Such transitions lead to splitting of the energy level of the
pair into a doublet. To second order in $J$ the energies of the
resulting symmetric and antisymmetric states are
\begin{equation}
\label{doublet}
E_{D}^{(\pm)} = E_D^{(0)} + \frac{J(2J\Delta +g)}{4\Delta (J\Delta +g)}
 \pm \frac{J^2}{4 (J\Delta + g)} .
\end{equation}
The energy splitting between the states is small if $J\Delta$ and $g$
have same sign. If on the other hand, $|J\Delta +g| \alt |J|$, the
theory has to be modified. Here, the bound pairs with one excitation
localized on the defect are resonantly mixed with extended unbound
two-magnon states. We do not consider this case in the present paper.

We note that, in terms of quantum computing, the onset of a doublet
suggests a simple way of creating entangled states. Indeed, by
applying a $\pi$ pulse at frequency $\varepsilon_1+g$ one can
selectively excite the qubit $n_0$. If then a $\pi$ pulse is applied
at the frequency $E_D^{+}-(\varepsilon_1+g)$, it will selectively
excite a Bell state $\left[|01\rangle + |10\rangle\right]/\sqrt{2}$,
where $|ij\rangle$ describes the state where the qubits on sites
$n_0-1$ and $n_0+1$ are in the states $|i\rangle$ and $|j\rangle$,
respectively. The excitations on sites $n_0\pm 1$ can then be
separated without breaking their entanglement using two-qubit
gate operations.

\subsubsection{Localized states split off the bound-pair band}

The energy difference $E_D^{(\pm)}-E_{BP}\approx g$ between BP states
on the defect site (\ref{doublet}) and extended BP states largely
exceeds the bandwidth $|J/\Delta|$ of the extended states. Therefore
it is a good approximation to assume that the wave functions of
extended BP states are equal to zero on the defect. In other words,
such BP's are reflected {\it before} they reach the defect. In this
sense, the defect acts as a boundary for them. One may expect that
there is a surface-type state associated with this boundary.

The emergence of the surface-type state is facilitated by the
defect-induced change of the on-site energy of a BP located next to
the defect on the sites $(n_0+1, n_0+2)$ [or $(n_0-2,n_0-1)$]. This
change arises because virtual dissociation of a BP with one excitation
hopping onto a defect site gives a different energy denominator
compared to the case where a virtual transition is made onto a regular
site. It is described by an extra term $\delta E_{BP}$ in the
expression (\ref{transfer_bound}) for the diagonal part of the
Hamiltonian $\tilde H_n^{(t)}$ with $n=n_0+1$ [or $n=n_0-2$],
\begin{equation}
\label{delta_BP}
\delta E_{BP}= \frac{Jg}{4\Delta(J\Delta - g)}.
\end{equation}

Using the transformed Hamiltonian (\ref{transferBP}), one can analyze
BP states in a way similar to the analysis of one-excitation states in
an open chain, see the Appendix. The Schr\"odinger equation for BP states
away from the defect is given by the first term in Eq.~(\ref{a(n,m)})
and by Eqs.~(\ref{transferBP}) and (\ref{delta_BP}). The BP wave functions
can be sought in the form
\begin{equation}
\label{BPfull}
a(n,n+1) = C e^{i\theta n} + C' e^{-i \theta n}.
\end{equation}
Then the BP energy as a function of $\theta$ is given by $E_{BP}=
E_{BP}^{(0)} + (J/2\Delta)\cos\theta$, cf. Eq.~(\ref{bound}).

The values of $\theta$ can be found from the boundary condition
that the BP wave function is equal to zero on the defect, i.e.,
$a(n_0,n_0+1) = a(n_0-1,n_0) \equiv a(n_0+N-1,n_0+N)= 0$. From the
Schr\"odinger equations for $a(n,n+1)$ with $n=n_0+1$ and $n=N+n_0-2$,
we obtain an equation for $\theta$ of the form
\begin{eqnarray}
\label{theta_BP}
f(\theta ) = f(-\theta ),\; f(\theta ) = \left[\delta E_{BP} -
{J\over 4 \Delta} e^{i \theta }\right]^2e^{i\theta (N-3)}.
\end{eqnarray}

Equation (\ref{theta_BP}) has $2(N-1)$ solutions for $\exp(i\theta)$. The
solutions $\exp(i\theta)=\pm 1$ are spurious, in the general case. The
roots $\theta$ and $-\theta$ describe one and the same wave
function. Therefore there are $N-2$ physically distinct roots
$\theta$, as expected for an $N$-spin chain with two excluded BP
states located at $(n_0, n_0\pm 1)$.

Depending on the ratio $q=J/(4\Delta\,\delta E_{BP})\equiv (J\Delta -
g)/g$, the roots $\theta$ are either all real or there is one or two
pairs of complex roots with opposite signs. Real roots correspond to
extended states, whereas complex roots correspond to the states that
decay away from the defect. The onset of complex roots can be analyzed
in the same way as described in the Appendix for one excitation. There is
much similarity, formally and physically, between the onset of
localized BP states next to the defect and the onset of surface states
at the edge of an open chain.

We
rewrite Eq.~(\ref{theta_BP}) in the form
\begin{equation}
\label{tangent_BP}
\tan[\theta(N-1)]={2\sin\theta \,(\cos \theta - q)\over (\cos\theta -
q)^2 - \sin^2\theta}.
\end{equation}
For $|q| > 1$ all roots of Eq.~(\ref{tangent_BP}) are real. At $|q|=1$
there occurs a bifurcation where two real roots with opposite signs
merge (at $\theta = 0$, for $q=1$, or at $\theta = \pi$, for
$q=-1$). They become complex for $|q| < 1$. For $|q| = 1-2(N-1)^{-1}$
two other real roots coalesce at $\theta = 0$ or $\pi$ and another
pair of complex roots emerges.

As the length of the chain increases, the difference between the pairs
of complex roots decreases. In the limit $N\to \infty$ the roots merge
pairwise. The imaginary part of one of the roots is%
\begin{equation}
\label{BP_s}
{\rm Im}~\theta_{BP}^{(s)}= \ln |\delta E_{BP}/(J/4\Delta )|.
\end{equation}
The second root has opposite sign.

One can show from the Schr\"odinger equation for the BP's that the
solution (\ref{BP_s}) describes a ``surface-type'' BP state localized
next to the defect on sites $(n_0+1, n_0+2)$. The solution with
$-\theta_{BP}^{(s)}$ describes the surface state on
$(n_0-2,n_0-1)$. The amplitudes of these states exponentially decay
away from the defect. For $J\Delta >g> J\Delta/2 > 0$ or $J\Delta <g<
J\Delta/2 < 0$ we have Re~$\theta_{BP}^{(s)}=0$.  For $g> J\Delta > 0$
or $g< J\Delta < 0$, we have Re~$\theta_{BP}^{(s)} =\pi$, and decay of
the localized state is accompanied by oscillations.  The complex roots
in a finite chain can be pictured as describing those same states on
the opposite sites of the defect. But now the states are
``tunnel'' split because of the overlap of their tails inside the
chain, which leads to the onset of two slightly different localization
lengths.

\begin{figure}[ht]
\includegraphics[width=2.8in]{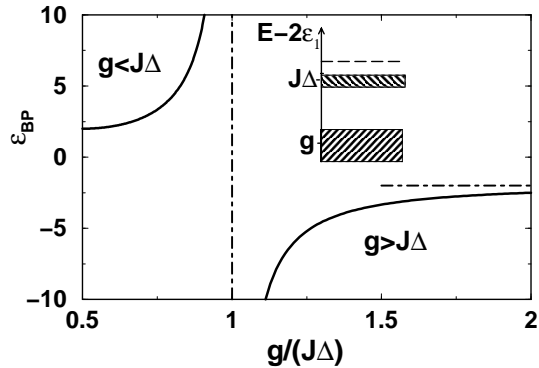}
\caption{The distance between the energy level of the BP localized
next to the defect and the center of the band of extended BP states,
$\varepsilon_{BP} = (E_{BP}^{(s)} - E_{BP}^{(0)})/(J/4\Delta) $, vs
the scaled defect excess energy $g/J\Delta$. The localized state
exists in an infinite chain for $g/J\Delta> 1/2$. The vertical and
horizontal dot-dashed lines show the asymptotes for $g\to J\Delta$ and
$g/J\Delta\to\infty$, respectively. The results refer to the range
where the BP and LDP bands are far from each other, compared to the
LDP bandwidth $|J|$. These bands are sketched in the inset (the BP band
is above the LDP band for $J\Delta/g > 1$). The dashed line in the
inset shows where the energy level of the localized BP state is
located with respect to the bands.}
\label{fig:BP_s}
\end{figure}

The energy of the localized state in a long chain is
\begin{equation}
\label{BP_s_energy}
E_{BP}^{(s)}= E_{BP}^{(0)}+ \delta E_{BP} +
\frac{(J/4\Delta )^2}{\delta E_{BP}}.
\end{equation}
It lies outside the band $ E_{BP}^{(0)}\pm (J/2\Delta)$ (\ref{bound})
of the extended BP states. The distance to the band edge strongly
depends on the interrelation between the defect excess energy $g$ and
the BP binding energy $J\Delta$. It is of the order of the BP
bandwidth $|J/\Delta|$, except for the range where the difference
between $J\Delta$ and $g$ becomes small. In this range the energy of
the surface-type state sharply increases in the absolute value. This
is illustrated in Fig.~\ref{fig:BP_s}. The localization length $|{\rm
Im}~\theta_{BP}^{(s)}|^{-1}$ is large when the state energy is close
to the band edge and shrinks down with decreasing $|J\Delta -g|$,
i.e., with increasing $|\delta E_{BP}|$.

\section{Antiresonant decoupling of two-excitation states}

The analysis of the preceding section does not apply if the pair
binding energy $J\Delta$ is close to the defect excess energy $g$.
When $|g-J\Delta|$ is of the order of the LDP bandwidth $|J|$, the BP
and LDP states are in resonance, their bands overlap or nearly overlap
with each other. One might expect that there would occur mixing of
states of these two bands. In other words, a delocalized magnon in the
LDP band might be scattered off the excitation on the defect, and as a
result they both would move away as a bound pair. However, as we show,
such mixing does not happen.  In order to simplify notations we will
assume in what follows that $g,J,\Delta >0$.

\subsection{A bound pair localized next to the defect}

In the resonant region we should reconsider the analysis of the
next-to-the-defect bound pair localized on sites $(n_0+1, n_0+2)$ or
$(n_0-2,n_0-1)$. As $|g-J\Delta|$ decreases, the energy of this pair
(\ref{BP_s_energy}) moves away from the BP band, see
Fig.~\ref{fig:BP_s}. At the same time, the distance between the pair
energy and the LDP band $E_{BP}^{(s)}- 2\varepsilon_1-g$ becomes {\it
smaller} with decreasing $|g-J\Delta|$ as long as $|J\Delta - g|>
|J|/2$. For $|J\Delta - g|\sim J$, the next-to-the-defect pairs are
hybridized with LDP's. The hybridization occurs in first order in the
nearest-neighbor coupling constant $J$.

To describe the hybridization, we will seek
the solution of the Schr\"odinger equation (\ref{a(n,m)}) in the form
of a linear superposition of an LDP state (\ref{LDP_wave}) and a
pair on the next to the defect sites,
\begin{eqnarray}
\label{hybridization}
a(n,m)&=& (Ce^{i\theta m} +C'e^{-i\theta m}) \delta_{n,n_0} \nonumber\\ &+&
a(n_0+1,n_0+2)  \delta_{n,n_0+1}\delta_{m,n_0+2} \nonumber\\
&+&
a(n_0-2,n_0-1) \delta_{n,n_0-2}\delta_{m,n_0-1}
\end{eqnarray}
(we remind that $m > n$). The energy of a state with given $\theta$
can be found from (\ref{a(n,m)}) as before by considering $m$ far away
from $n_0$. It is given by Eq.~(\ref{energy_LDP}).

The interrelation between the coefficients $C, C', a(n_0+ 1, n_0+2)$,
and $a(n_0-2, n_0-1) $, as well as the values of $\theta$ should be
obtained from the boundary conditions. These conditions follow from
the fact that the energy of a pair on sites $(n_0, n_0\pm 1)$ and the
energies of the pairs described by Eq.(\ref{hybridization}) differ by
$\sim g\approx J\Delta$, cf. (\ref{doublet}). Therefore the pairs on
sites $(n_0, n_0\pm 1)$ are decoupled from the states
(\ref{hybridization}), and in the analysis of the LDP's we can set
$a(n_0, n_0\pm 1)=0$. Decoupled also are unbound two-excitation states
with no excitation on the defect, i.e., two-magnon states. Therefore
$a(n,m)=0$ if simultaneously $m-n>1$ and $(n-n_0)(m-n_0)\neq 0$.

From Eq.~(\ref{a(n,m)}) written for $n=n_0,m=n_0+ 2$ and
$n=n_0-2,m=n_0$ [with account taken of the relation $a(n,m)=a(m,n+N)$]
we obtain
\begin{eqnarray}
\label{hybr_interrelation}
a(n_0+1, n_0+2)&=& Ce^{i\theta (n_0+1)} +C'e^{-i\theta (n_0+1)},\\
a(n_0-2, n_0-1)&=& Ce^{i\theta (n_0+N-1)} +C'e^{-i\theta (n_0+N-1)}\nonumber .
\end{eqnarray}

The interrelation between $C$ and $C'$ and the equation for $\theta$
follow from Eqs.~(\ref{hybridization}), (\ref{hybr_interrelation}) 
and (\ref{a(n,m)}) written
for $n=n_0+1,m=n_0+ 2$ and $n=n_0-2,m=n_0-1$. They have the form
\begin{equation}
\label{hybr_cc'}
C' = -\frac{2(J \Delta -g) e^{i\theta } -J}
{2(J \Delta-g) e^{-i\theta } -J} Ce^{2i\theta n_0},
\end{equation}
and
\begin{eqnarray}
\label{hybr_theta}
&\tilde f(\theta)=\tilde f(-\theta),\\
&\tilde f(\theta)=e^{i \theta N} \left[2(J \Delta-g)e^{-i\theta} -
J \right]^2.\nonumber
\end{eqnarray}

Equation (\ref{hybr_theta}) is an $2N$th order equation for
$\exp(i\theta)$. Its analysis is completely analogous to that of
Eq.~(\ref{theta_BP}). The roots $\theta=0,\pi$ are spurious, and the
roots $\theta$ of opposite signs describe one and the same wave
function. Therefore Eq.~(\ref{hybr_theta}) has $N-1$ physically
distinct roots $\theta$. Real $\theta$'s correspond to extended
states. Complex roots appear for $2|J \Delta-g|
>J$. These roots, $\theta_{LDP}^{(s)}$, describe localized states. In
the limit of a long chain, $N\to \infty$,
the imaginary part of one of them is
\begin{equation}
\label{hybr_decrem}
{\rm Im}\, \theta_{LDP}^{(s)} =  \ln |2(J\Delta - g)/J|,
\end{equation}
whereas the other root has just opposite sign.

The wave function of the localized state is maximal either on sites
$(n_0+1, n_0+2)$ or $(n_0-2,n_0-1)$ and exponentially decays into the
chain. For $J\Delta - g < 0$ this decay is accompanied by
oscillations, Re~$\theta_{LDP}^{(s)}=\pi$.  The energy of the
localized state is
\begin{equation}
\label{hybr_s_energy}
E_{LDP}^{(s)} =  2{\varepsilon}_1 + J \Delta  + \frac{(J/2)^2}{J\Delta - g}.
\end{equation}

The localized state (\ref{hybridization}), (\ref{hybr_decrem}) is a
``surface-type'' state induced by the defect. It is the resonant-region
analog of the localized next-to-the-defect state discussed in
Section~III.B3. The wave function of the latter state (\ref{BPfull}),
(\ref{BP_s}) was a linear combination of the wave functions of bound
pairs. In contrast, the state given by Eqs.~(\ref{hybridization}),
(\ref{hybr_decrem}) is a combination of the wave functions of the bound
pair located next to the defect and a localized-delocalized pair.

The evolution of the surface-type state is controlled by the
difference between the excess energies of binding two excitations in a
pair or localizing one of them on the defect $|J\Delta - g|$. As
$|J\Delta - g|$ varies, the state changes in the following way. It
first splits off the BP band when $|J\Delta - g|$ becomes less than
$g$, see Fig.~\ref{fig:BP_s}. Its energy moves away from the band of
extended BP states with decreasing $|J\Delta - g|$ and the
localization length decreases [cf. Eq.~(\ref{BP_s})]. Well before
$|J\Delta - g|$ becomes of order $J$, the state becomes strongly
localized on the sites $(n_0+1, n_0+2)$ or $(n_0-2, n_0-1)$.

\begin{figure}[ht]
\includegraphics[width=2.8in]{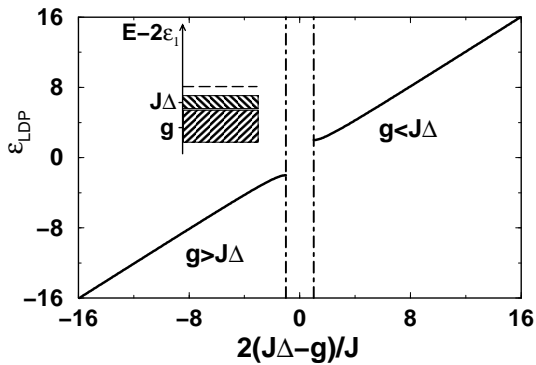}
\caption{The energy difference between the localized BP state and the
LDP band, $\varepsilon_{LDP} = (E_{LDP}^{(s)} - 2{\varepsilon}_1 -
g)/(J/2)$, vs the energy mismatch $2(J\Delta -g)/J$.
In the region between the dot-dashed lines the BP state is
delocalized. The results refer to the case where the BP and LDP bands
are close to each other or overlap, as sketched in the inset (the BP band
is above the LDP band for $J\Delta > g$).  The dashed line in the
inset shows where the energy level of the localized BP
state is located with respect to the bands.}
\label{fig:LDP}
\end{figure}

In the region $|J\Delta - g|\sim J$ the localized state becomes
stronger hybridized with LDP states than with extended BP states. This
hybridization occurs in first order in $J$, via a transition
$(n_0+1,n_0+2) \to (n_0, n_0+2)$ [or $(n_0-2,n_0-1) \to
(n_0-2,n_0)$]. In this region the localization length increases with
decreasing $|J\Delta - g|$, cf. Eq.~(\ref{hybr_decrem}). Ultimately,
for $|J\Delta - g|=J/2$ the localized surface-type state disappears,
as seen in Fig.~\ref{fig:LDP}. The evolution of the energy of the
localized state with $J\Delta - g$ is shown in Fig.~\ref{fig:LDP}.

The crossover from hybridization of the localized surface-type BP
state with extended BP states to that with LDP states occurs in the
region $|g|\gg |J\Delta - g| \gg |J|$. It is described using a
different approach in Ref.~\onlinecite{us_DS}. We note that the
expressions for the energy of the localized states (\ref{BP_s_energy})
and (\ref{hybr_s_energy}) go over into each other for $|J\Delta -
g|\ll |g|$.

\subsection{Decoupling of  bound pairs and LDP's
in the resonant region}

We are now in a position to consider resonant coupling between
extended states of bound pairs and localized-delocalized pairs. To
lowest order in $\Delta^{-1}$, a BP on sites $(n, n+1)$ far away from
the defect can resonantly hop only to the nearest pair of sites, see
Eqs.~(\ref{transfer_bound}), (\ref{transferBP}). As noted before, the
hopping requires an intermediate virtual transition of the BP into a
dissociated state, which differs in energy by $ J\Delta$.

A different situation occurs for the BP on the sites $(n_0+2,n_0+3)$
[or $(n_0-3,n_0-2)$]. Such BP can hop onto the sites $(n_0+1,n_0+2)$
and $(n_0+3,n_0+4)$, as described by (\ref{transferBP}). But in
addition, for $|J\Delta - g|\alt J$ it can make a
transition into the LDP state on sites $(n_0,n_0+3)$
[or $(n_0-3,n_0)$] . Indeed, such
state has the energy $\approx 2{ \varepsilon}_1+g$, which is
close to the BP energy $\approx 2{ \varepsilon}_1+J\Delta$.

The transition $(n_0+2,n_0+3) \to (n_0,n_0+3)$ goes through the
intermediate dissociated state $(n_0+1,n_0+3)$, which differs in
energy by $\approx J\Delta$.  It can be taken into account by adding
the term $\delta\tilde H^{(t)}$ to the BP hopping Hamiltonian
(\ref{transferBP}) for the sites $(n_0+2,n_0+3)$,
\begin{equation}
\label{extra_H_t}
\delta \tilde H^{(t)}a(n_0+2,n_0+3) = (J/4\Delta)a(n_0,n_0+3).
\end{equation}

Extended BP states are connected to LDP states only through a BP on
the sites $(n_0+2,n_0+3)$. We are now in a position to analyze this
connection.  From Eqs.~(\ref{transferBP}) and (\ref{extra_H_t}), we see
that
\begin{eqnarray}
\label{coupling_resonant}
&&[\tilde H^{(t)} + \delta \tilde H^{(t)}]a(n_0+2,n_0+3) =
(J/4\Delta)\nonumber\\
&&\times[a(n_0,n_0+3) + a(n_0+1,n_0+2)] + A,
\end{eqnarray}
where $A$ is a linear combination of the amplitudes $a(n_0+2,n_0+3)$
and $a(n_0+3,n_0+4)$.

The sum $a(n_0,n_0+3)+ a(n_0+1,n_0+2)$ in
Eq.~(\ref{coupling_resonant}) can be expressed in terms of the LDP
wave functions (\ref{hybridization}). With account taken of the
interrelation (\ref{hybr_cc'}) between the coefficients $C,C'$ in
Eq.~(\ref{hybridization}), we have
\begin{eqnarray}
\label{amplitude_resonant}
&&a(n_0+1,n_0+2)+a(n_0,n_0+3)\nonumber\\
&&\propto C \sin \theta \cos \theta [J\Delta -g -
J \cos \theta ].
\end{eqnarray}
Equations (\ref{coupling_resonant}), (\ref{amplitude_resonant}) describe
the coupling between the BP on the sites $(n_0+2,n_0+3)$ and the LDP
eigenstates with given $\theta$.

An important conclusion can now be drawn regarding the behavior of BP
and LDP states in the resonant region. The center of the BP band lies
at $2{\varepsilon}_1 + J\Delta +(J/2\Delta)$ (\ref{bound}), and the BP
band is parametrically narrower than the LDP band $2{\varepsilon}_1+g
+J\cos\theta$ (\ref{energy_LDP}).  When the BP band is inside the LDP
band, this means that, for an appropriate wave number $\theta$ of the
LDP magnon, $J\Delta = g+J\cos\theta$, to zeroth order in
$\Delta^{-1}$ (this is the approximation used to obtain the LDP
dispersion law). It follows from Eqs.~(\ref{coupling_resonant}) and
(\ref{amplitude_resonant}) that, for such $J\Delta -g$ and $\theta$
there is no coupling between the LDP and extended BP states.

The above result means that LDP and extended BP states do not experience
resonant scattering into each other, even though it is allowed by the
energy-conservation law. Such scattering would correspond to the
scattering of a magnon off the excitation localized on the
defect, with both of them becoming a bound pair that moves away from
the defect, or an inverse process.

Physically, the antiresonant decoupling of BP and LDP states is a
result of strong mixing of a BP on the sites $(n_0+1,n_0+2)$ and
LDP's. Because of the mixing, the amplitudes of transitions of extended
BP states to the sites $(n_0+1,n_0+2)$ and $(n_0,n_0+3)$ compensate
each other, to lowest order in $\Delta^{-1}$.

\begin{figure}[ht]
\includegraphics[width=3.2in]{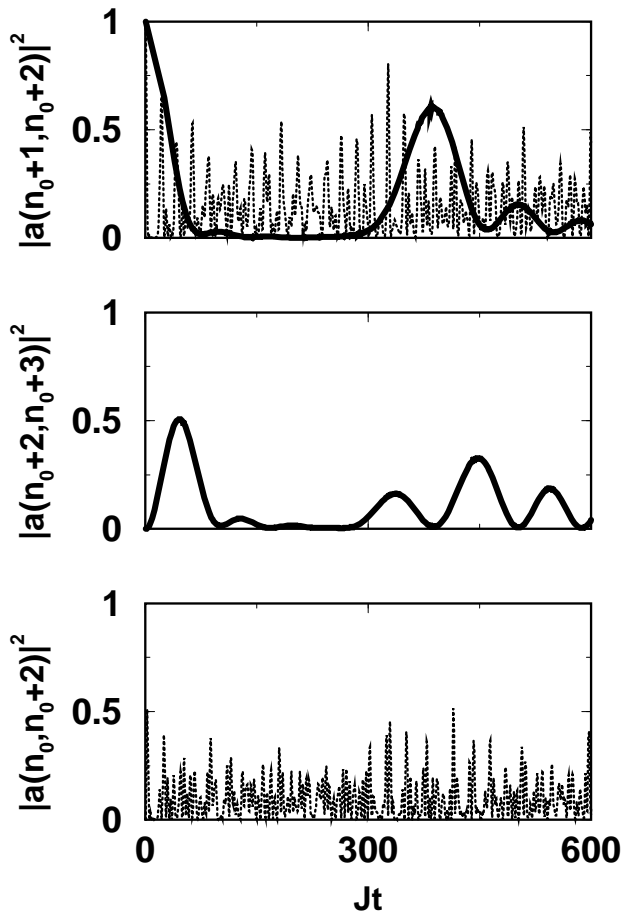}
\caption{Time evolution of a two-excitation wave packet in an $XXZ$
chain with a defect; $|a(n,m)|^2$ is the occupation of sites
$(n,m)$. Initially an excitation pair is placed next to the defect on
sites $(n_0+1,n_0+2)$. The solid and dashed lines refer to the cases
where the BP and LDP bands are, respectively, far away from each other
($g=J\Delta/4$) and overlapping ($g=J\Delta$). In the first case a
bound pair slowly oscillates between neighboring sites $(n,n+1)$ and
does not dissociate [$a(n_0,n_0+2)$ remains very small]. In the case
of overlapping bands, the pair at $(n_0+1,n_0+2)$ is hybridized with
LDP's, but practically does not mix with bound pairs on other sites
[$a(n_0+2,n_0+3)$ remains very small]. The results refer to a ten-site
closed chain with $\Delta =10$.}
\label{fig:temporal_nonresonant}
\end{figure}

To illustrate the antiresonant decoupling we show in
Fig.~\ref{fig:temporal_nonresonant} two types of time evolution of
excitation pairs. In both cases the initial state of the system was
chosen as a pair on the sites $(n_0+1,n_0+2)$, i.e., $a(n_0+1,n_0+2)=1$
for $t=0$. The time-dependent Schr\"odinger equation was then solved
with the boundary condition that corresponds to a closed chain.

The solid lines refer to the case of nonoverlapping BP and LDP bands,
$|J|\ll |g-J\Delta|$ and $|g|<|J\Delta|/2$.  In this case there does
not emerge a localized surface-type BP state next to the
defect. Therefore an excitation pair placed initially on the sites
$(n_0+1,n_0+2)$ resonantly transforms into extended BP states and
propagates through the chain. This propagation is seen from the figure
as oscillations of the return probability and the probability to find
the BP on another arbitrarily chosen pair of neighboring sites
$(n_0+2,n_0+3)$.  The oscillation rate should be small, of the order
of the bandwidth $J/\Delta$. This estimate agrees with the numerical
data. It is seen that the BP state is not transformed into LDP
states. The amplitude of LDP states on sites $(n_0,m\geq n_0+2)$
remains extremely small, as illustrated for $m=n_0+2$.

The dotted lines in Fig.~\ref{fig:temporal_nonresonant} show a
completely different picture which arises when the BP band is inside
the LDP band. In this case an excitation pair placed initially on
$(n_0+1,n_0+2)$ hybridizes with LDP rather than BP states. A
transformation of the pair $(n_0+1,n_0+2)$ into LDP's with increasing
time is clearly seen. The period of oscillations is of the order of
the reciprocal bandwidth of the LDP's $J^{-1}$, it is much shorter than in
the previous case.  Remarkably, as a consequence of the antiresonance,
extended states of bound pairs are not excited to any appreciable
extent, as seen from the amplitude of the pair on the sites
$(n_0+2,n_0+3)$.

\section{Conclusions}

We have analyzed the dynamics of a disordered spin chain with a
strongly anisotropic coupling in a magnetic field. A defect in such a
chain can lead to several localized states, depending on the number of
excitations. This is a consequence of the interaction between
excitations and its interplay with the disorder. We have studied
chains with one and two excitations.

The major results refer to the case of two excitations. Here, the
physics is determined by the interrelation between the excess on-site
energy of the defect $g$ and the anisotropic part of the exchange
coupling $J\Delta$.  Strong anisotropy leads to binding of excitations
into nearest-neighbor pairs that freely propagate in an ideal
chain. Because of the defect, BP's can localize. A simple
type of a localized BP is a pair with one of the excitations located
on a defect.  A less obvious localized state corresponds to a pair
localized next to a defect. It reminds of a surface state split off
from the band of extended BP states, with the surface being the defect
site.  We specified the conditions where the localization occurs and
found the characteristics of the localized states.

Our most unexpected observation is the antiresonant decoupling of 
extended BP states from localized-delocalized pairs. The LDP's are
formed by one excitation on the defect site and another in an extended
state. The antiresonance occurs for $g\approx J\Delta$, when the BP and
LDP bands overlap. It results from destructive quantum interference of
the amplitudes of transitions of BP's into two types of resonant
two-excitation states: one is an LDP, and the other is an excitation
pair on the sites next to the defect. As a result of the
antiresonant decoupling, extended BP's and LDP's do not \
scatter into each other, even though the
scattering is allowed by energy conservation. This means that an
excitation localized on the defect does not delocalize as a result of
coupling to other excitations.

The occurrence of multiple localized states in the presence of other
excitations is important for quantum computing.  It shows that, even
where the interaction between the qubits is ``on'' all the time, we
may still have well-defined states of individual qubits that can be
addressed and controlled. One can prepare entangled localized pairs of
excitations, as we discussed in Sec.~III~B~2, or more complicated
entangled excitation complexes.  The results of the paper also provide
an example of new many-body effects that can be studied using quantum
computers with individually controlled qubit transition energies.

\begin{acknowledgments}
This research was supported in part by the NSF through Grant
No. ITR-0085922 and by the Institute for Quantum Sciences at Michigan
State University.
\end{acknowledgments}

\appendix

\section{One excitation}

In an infinite spin chain in a magnetic field, the anisotropy of the
spin-spin interaction does not affect the spectrum and wave functions
of one excitation.  The matrix element of the term
$\sum_n\sigma_n^z\sigma_{n+1}^z$ in the Hamiltonian
(\ref{hamiltonian}) is just a constant. However, the situation becomes
different for a chain of finite length, because the coupling
anisotropy can lead to surface states. In the case of two excitations,
analogs of surface-type states emerge near defects 
in an infinite chain, as discussed in Sec.~III. 
Here, for completeness and keeping
in mind a reader with the background in quantum computing, we briefly
outline the results of the standard analysis of a finite-length spin
chain with one excitation.

\subsection{Localized surface and defect-induced states in an open chain}

The Schr\"odinger equation for an excitation in an open chain has the
form (\ref{eq_a}), and its solution $a(n)$ to the left and to the
right from the defect can be written in the form of a superposition of
counterpropagating plane waves, Eq.~(\ref{left_right}). The relation
between the amplitudes of these waves $C'_{l,r}$ and $C_{l,r}$ follows
from the boundary conditions $a(0)=a(N+1)=0$. By substituting
Eq.~(\ref{left_right}) into Eq.~(\ref{eq_a}) with $n=1$ and $n=N$, we
obtain
\begin{eqnarray}
\label{bound_cond}
&&C'_l = - D C_l, \quad C'_r = -D^{-1} e^{2i\theta (N+1)} C_r,\\
&&D = [1-\Delta \exp (i\theta )]/[1-\Delta \exp (-i \theta )].\nonumber
\end{eqnarray}

The relations between $C'_{l,r}$, $C_{l,r}$, and the amplitude of the
wave function on the defect site $a(n_0)$ follow from
Eqs.~(\ref{eq_a}) and (\ref{left_right}) for $n=n_0\pm 1$,
\begin{eqnarray}
\label{boundary_defect}
a(n_0)&&= C_{l} e^{i\theta n_0}\left[1 - D e^{-2i\theta n_0}\right]\nonumber\\
&&= C_{r} e^{i\theta n_0}\left[1 - D^{-1}
e^{2i\theta (N+1-n_0)}\right] .
\end{eqnarray}
With (\ref{bound_cond}) and (\ref{boundary_defect}), all coefficients in
the wave function (\ref{left_right}) are expressed in terms of one
number, $a(n_0)$. It can be obtained from normalization.

In a finite chain, the values of $\theta$ are quantized. They can be
found from Eq.~(\ref{eq_a}) with $n=n_0$. With account taken of
(\ref{boundary_defect}), this equation can be written as
\begin{eqnarray}
\label{theta}
f_N (\theta ) - D^2 f_N (-\theta ) = -(ig/J\sin \theta )
\qquad\qquad\qquad\nonumber\\
 \times [
f_N (\theta ) + D^2 f_N (-\theta ) - 2D
\cos \theta (N-2 n_0 +1) ],
\end{eqnarray}
where
\begin{equation}
f_N (\theta ) = \exp[i\theta (N+1)] .
\end{equation}

The analysis of the roots of Eq.~(\ref{theta}) is standard. This is a
$2(N+1)$-order equation for $\exp(i\theta)$, but its solutions for
$\theta$ come in pairs $\theta$ and $-\theta$. Each pair gives one
wave function, as seen from Eq.~(\ref{left_right}). In addition,
Eq.~(\ref{theta}) has roots $\theta = 0,\pi$; they are spurious
(unless $|\Delta| = 1$ and $g= 0$) and appear as a result of algebraic
transformations. Therefore Eq.~(\ref{theta}) has $N$ physically
distinct roots, as expected for a chain of $N$ spins.  We note that,
for $g=0$ the position of the impurity $n_0$ drops out from
Eq.~(\ref{theta}), and then the equation goes over into the result for
an ideal chain.

Solutions of Eq.~(\ref{theta}) with real $\theta$ correspond, in the
case of a long chain, to delocalized magnon-type excitations
propagating in the chain. Their bandwidth is $2|J|$.

Along with delocalized states, Eq.~(\ref{theta}) describes also
localized states with complex $\theta$. Complex roots of
Eq.~(\ref{theta}) can be found for a long chain, where $|{\rm
Im}\,\theta|(N-n_0),\, |{\rm Im}\,\theta|n_0 \gg 1$. They describe
surface states, which are localized on the chain boundaries, and a
state localized on the defect. The localization length of the states
is given by $|1/{\rm Im}\,\theta|$.

The surface states arise only for the
anisotropy parameter $|\Delta|>1$. The corresponding values of
$\theta$ are
\begin{equation}
\label{bound_local}
\theta_s = \pm i\ln|\Delta| + \pi\Theta(-\Delta).
\end{equation}
Here, the signs $+$ and $-$ refer to the states localized on the
left and right boundaries, respectively, and $\Theta(x)$ is the step
function.

From Eq.~(\ref{energy_one}), the energy of the surface state is
\begin{equation}
\label{E_s}
E_s={\varepsilon}_1 + J(\Delta^2 +1)/2\Delta
\end{equation}
It lies outside the energy band of delocalized excitations. We note
that, for $\Delta > 1$ the surface states decay monotonically with the
distance from the boundary (Re~$\theta=0$). For sufficiently large
negative $\Delta$, on the other hand, the decay of the wave function
is accompanied by oscillations, and $a(n)$ changes sign from site to site.

A defect in a long chain gives rise to a localized one-spin excitation
for an arbitrary excess energy $g$ \cite{Economou}. The amplitude
$a(n)$ decays away from the defect as
\begin{eqnarray}
\label{one_localized}
&a(n)=a(n_0)\exp(i\theta_d|n-n_0|),\\
&\theta_d=i\sinh^{-1}(|g/J|)+
\pi\Theta(-g/J).\nonumber
\end{eqnarray}
The energy of the localized state is
\begin{equation}
\label{E_d}
E_d=\varepsilon_1+
(g^2+J^2)^{1/2}{\rm sgn}\,g.
\end{equation}

For small $|g/J|$, we have Im~$\theta_d \approx |g/J|$, i.e., the
reciprocal localization length is simply proportional to the defect
excess energy $|g|$. In the opposite case of large $|g/J|$ we have
Im~$\theta_d = \ln |2g/J|$. In this case the amplitude of the
localized state rapidly falls off with the distance from the defect,
$a(n)\propto (2|g/J|)^{-|n-n_0|}$.

\begin{figure}[ht]
\includegraphics[width=3.2in]{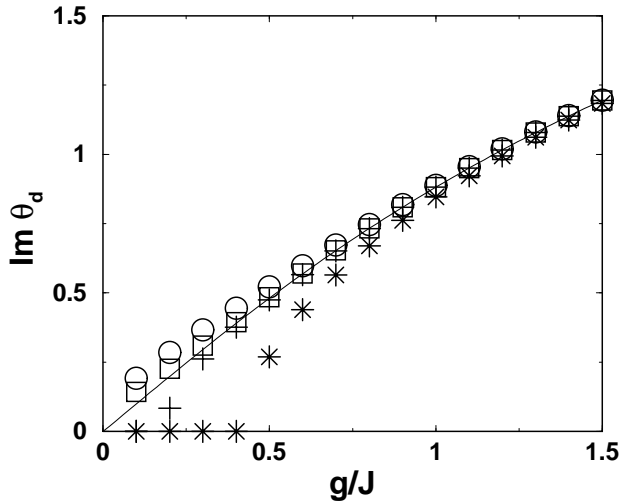}
\caption{ The reciprocal localization length Im~$\theta_d$ for an
infinite chain [Eq.~(\protect\ref{one_localized})] as a function of
the defect excess energy $g$ for $J=1$ (solid line). Also shown are
the results for an open chain with
the number of sites $N=6$ and 12 and $n_0=N/2$ (stars and crosses,
respectively) and a closed chain with 6 and 12 sites (circles and
squares, respectively). They are obtained from
Eq.~(\protect\ref{theta}) with $\Delta = 10$ and from
Eq.~(\protect\ref{dispersion}). In an open chain, solutions with
nonzero Im~$\theta_d$ emerge starting with a certain
$|g/J|>|g/J|_{\min}$. In a closed chain with even $N$ there is no
threshold in $|g/J|$ for the onset of states with Im~$\theta_d\neq
0$.}
\label{fig:loc_length}
\end{figure}

If the localization length is comparable to the chain length, the
notion of localization is not well defined. However,
when discussing numerical results, one can formally call
a state localized if its wave function exponentially decays away from
the defect and is described by a solution of Eq.~(\ref{theta}) with
complex $\theta$. This is equivalent to the statement that the state
energy $E_d$ lies outside the band of magnons in the infinite
chain. In an open finite chain such localized state may emerge
provided the localization length is smaller than the distance from the
defect to the boundaries. This means that the defect excess energy
$|g|$ should exceed a minimal value that depends on the size of the
chain. The comparison of Eq.~(\ref{one_localized}) with the numerical
solutions of the full equation (\ref{theta}) for a finite chain is shown in
Fig.~\ref{fig:loc_length}.

\subsection{One-excitation states in a closed chain}

As pointed out in Sec.~II, for a closed chain the solution of the
Schr\"odinger equation can be also sought in the form of
counterpropagating waves with different amplitudes,
Eq.~(\ref{right}). Clearly, the phases $\theta$ and $-\theta$ describe
one and the same wave function.  The one-excitation energy $E_1$ is
given by Eq.~(\ref{energy_one}).

The interrelation between the amplitudes of the waves $C, C'$ in
Eq.~(\ref{right}) and the amplitude of the wave function on the defect
site $a(n_0)$ can be obtained from Eq.~(\ref{eq_a}) with $n=n_0\pm
1$. This equation has two solutions,
\begin{equation}
\label{untouched}
e^{i\theta N}=1,\quad a(n_0)=Ce^{i\theta n_0}+C'e^{-i\theta n_0}
\end{equation}
and
\begin{eqnarray}
\label{touched}
a(n_0)&=&Ce^{i\theta n_0}\left(1+e^{i\theta N}\right)\\
&=&C'e^{-i\theta n_0}\left(1+e^{-i\theta N}\right)\quad [\exp(i\theta
N)\neq 1].\nonumber
\end{eqnarray}
In order to fully determine the wave function $a(n)$ (\ref{right}),
Eqs.~(\ref{untouched}), (\ref{touched}) should be substituted into the
Schr\"odinger equation (\ref{eq_a}) for $n=n_0$.

Equations (\ref{eq_a}) and (\ref{untouched}) can be satisfied provided that
either $g=0$, which means that there is no defect, or $a(n_0)=0$. The
first condition describes excitations in an ideal closed chain and is
not interesting for the present paper. The condition $a(n_0)=0$
corresponds to the wave function $a(n)\propto \sin \theta (n-n_0)$,
which has a simple physical meaning. It is a standing wave in an ideal
chain with a node at the location of a defect. Because of the node,
the corresponding state ``does not know'' about the defect, and
therefore it is exactly the same as in an ideal chain.

In the presence of a defect, the solutions of Eq.~(\ref{untouched}) in
the range of interest $0<\theta < \pi$ are $\theta = 2\pi k/N$ with
$k=1,2,\ldots,(N-1)/2$ for odd $N$, or $k=1,2,\ldots,N/2-1$ for even
$N$.

The equation for $\theta$ that follows from Eqs.~(\ref{eq_a}) and
(\ref{touched}) has the form
\begin{equation}
\label{dispersion}
\exp (i\theta N) - 1 = -\frac{ig}{J \sin \theta} [\exp (i\theta N)+1] .
\end{equation}
For $g\neq 0$ this equation has either $(N+1)/2$ (for odd $N$) or
$N/2+1$ (for even $N$) solutions for $\pm \theta$. Therefore the total
number of solutions for $\theta$ that follow from
Eqs.~(\ref{untouched}) and (\ref{dispersion}) is $N$, as expected.

By rewriting Eq.~(\ref{dispersion}) as $\tan(\theta N/2) =
-(g/J\sin\theta)$ and plotting the left- and right-hand sides as
functions of $\theta$ (cf. Ref.~\onlinecite{Economou}), one can see
that all physically distinct roots of this equation but one are real
and lie in the interval $0< \theta < \pi$ [except for one case, see
below]. Such solutions describe delocalized states with
sinusoidal wave functions.

The complex root of Eq.~(\ref{dispersion}), $\theta=\theta_d$,
describes a state localized on the defect. For a long chain,
Im~$\theta_d N\gg 1$, the solution has the form (\ref{one_localized}),
as expected. An interesting situation occurs for a shorter chain. If
$N$ is even or if $g/J>0$, a localized solution with complex
$\theta_d$ emerges for any defect excess energy $g$.  Thresholdless
localization does not happen in an open chain. In a closed chain, it
arises because there is no reflection from boundaries. For small
positive $g/J$ one obtains the complex solution of
Eq.~(\ref{dispersion}) in the form $\theta_d\approx i(2g/NJ)^{1/2}$.
The square-root dependence of Im~$\theta_d$ on $g$ is seen from
Fig.~\ref{fig:loc_length}.

Equation (\ref{dispersion}) has a complex
solution also for even $N$ and small negative $g/J$.  In this case
$\theta_d\approx i(-2g/NJ)^{1/2} + \pi$, i.e., the decay of the wave
function $a(n)$ is accompanied by sign flips, $a(n+1)/a(n) <
0$. Such oscillations cannot be reconciled with the
periodicity condition for odd $N$. Therefore, for odd $N$ and negative
$g/J$ a decaying solution arises only when $-g/J$ exceeds a threshold
value. One can show from Eq.~(\ref{dispersion}) that this value is
$|g/J|_{\min}=2/N$, which has also been confirmed numerically.

\end{document}